\documentclass[aps,prl,twocolumn,superscriptaddress, showkeys]{revtex4-1}

\usepackage{graphicx}
\usepackage{siunitx}
\usepackage{empheq}
\usepackage{amsmath}
\usepackage{amssymb}
\usepackage{txfonts}
\usepackage[colorlinks=true, linkcolor=blue,citecolor=red]{hyperref}
\usepackage{xcolor}
\usepackage{ulem,xpatch}

\renewcommand{\Re}{\operatorname{Re}}
\renewcommand{\Im}{\operatorname{Im}}
\renewcommand{\figurename}{Figure}
\newcommand{\im}{\mathrm{i}}
\newcommand{\e}{\mathrm{e}}

\newcommand{\rp}[1]{(\ref{#1})}

\newcommand{\abs}[1]{\left|{#1}\right|}

\newcommand{\av}[1]{\left\langle #1 \right\rangle}

\newcommand{\wt}[0]{\widetilde}

\newcommand{\da}{^\dagger}

\newcommand{\pt}[1]{\left( #1 \right)}
\newcommand{\pq}[1]{\left[ #1 \right]}
\newcommand{\pg}[1]{\left\{ #1 \right\}}

\newcommand{\lpg}[1]{\left\{ #1 \right.}

\newcommand{\rpg}[1]{\left. #1 \right\}}
\newcommand{\ee}{{\rm e}}
\newcommand{\ii}{{\rm i}}
\newcommand{\dd}{{\rm d}}

\newcommand{\nn}{{\nonumber}}

\newcommand{\EE}{{\cal E}}

\newcommand{\II}{{\cal I}}

\newcommand{\PP}{{\cal P}}

\newcommand{\ZZ}{{\cal Z}}

\renewcommand{\Re}{\operatorname{Re}}
\renewcommand{\Im}{\operatorname{Im}}
\renewcommand{\figurename}{Fig.}
\newcommand{\eff}{\mathrm{eff}}
\newcommand{\fb}{\mathrm{fb}}
\newcommand{\mcg}{\mathcal{G}}

\newcommand{\wc}{\omega_\mathrm{c}}
\newcommand{\ko}{\kappa_0}
\newcommand{\kp}{\kappa^{\prime}}
\newcommand{\kpp}{\kappa^{\prime\prime}}
\newcommand{\wm}{\omega_\mathrm{m}}
\newcommand{\gm}{\gamma_\mathrm{m}}
\newcommand{\dfq}{\dot q}
\newcommand{\dfp}{\dot p}
\newcommand{\dfa}{\dot a}
\newcommand{\dta}{\delta a}
\newcommand{\dtq}{\delta q}

\newcommand{\gfb}{g_\mathrm{fb}}
\newcommand{\tgfb}{\tilde g_\mathrm{fb}}
\newcommand{\ain}{a_\mathrm{in}}
\newcommand{\ainp}{a_\mathrm{in}^\prime}
\newcommand{\ainpp}{a_\mathrm{in}^{\prime\prime}}

\newcommand{\tcfb}{\tilde\chi_\mathrm{fb}}
\newcommand{\as}{\alpha_\mathrm{s}}
\newcommand{\tcc}{\tilde\chi_\mathrm{c}}
\newcommand{\tccs}{\tilde\chi_\mathrm{c}^\ast}

\newcommand{\tcce}{\tilde\chi_\mathrm{c}^\mathrm{eff}}

\newcommand{\tcm}{\tilde\chi_\mathrm{m}}

\newcommand{\tcmoe}{\tilde\chi_\mathrm{m}^\mathrm{o,eff}}

\begin{document}

\title{Normal--mode splitting in a weakly coupled optomechanical system}
\author{Massimiliano Rossi}
	\altaffiliation[Present address: ]{\footnotesize \textit{Niels Bohr Institute, University of Copenhagen, Blegdamsvej 17, 2100 Copenhagen, Denmark}}
	\affiliation{School of Higher Studies ``C. Urbani", University of Camerino, 62032 Camerino (MC), Italy}
	\affiliation{School of Science and Technology, Physics Division, University of Camerino, 62032 Camerino (MC), Italy}
\author{Nenad Kralj}
	\affiliation{School of Science and Technology, Physics Division, University of Camerino, 62032 Camerino (MC), Italy}
\author{Stefano Zippilli}
\author{Riccardo Natali}
	\affiliation{School of Science and Technology, Physics Division, University of Camerino, 62032 Camerino (MC), Italy}
	\affiliation{INFN, Sezione di Perugia, 06123 Perugia (PG), Italy}
\author{Antonio Borrielli}
	\affiliation{Institute of Materials for Electronics and Magnetism, Nanoscience-Trento-FBK Division, 38123 Povo (TN), Italy}
\author{Gregory Pandraud}
	\affiliation{Delft University of Technology, Else Kooi Laboratory, 2628 Delft, The Netherlands}
\author{Enrico Serra}
	\affiliation{Delft University of Technology, Else Kooi Laboratory, 2628 Delft, The Netherlands}
	\affiliation{Istituto Nazionale di Fisica Nucleare, TIFPA, 38123 Povo (TN), Italy}
\author{Giovanni Di Giuseppe}
	\email{gianni.digiuseppe@unicam.it}
	\affiliation{School of Science and Technology, Physics Division, University of Camerino, 62032 Camerino (MC), Italy}
	\affiliation{INFN, Sezione di Perugia, 06123 Perugia (PG), Italy}
\author{David Vitali}
	\email{david.vitali@unicam.it}
	\affiliation{School of Science and Technology, Physics Division, University of Camerino, 62032 Camerino (MC), Italy}
	\affiliation{INFN, Sezione di Perugia, 06123 Perugia (PG), Italy}
	\affiliation{CNR-INO, L.go Enrico Fermi 6, I-50125 Firenze, Italy}
\date{\today}

\begin{abstract}
Normal--mode splitting is the most evident signature of strong coupling between two interacting subsystems. 
It occurs when two subsystems exchange energy between themselves faster than they dissipate it to the environment. Here we experimentally show
that a weakly coupled optomechanical system at room temperature can manifest normal--mode splitting when the pump field fluctuations are anti-squashed by a phase-sensitive feedback loop operating close to its instability threshold. Under these conditions the optical cavity exhibits an effectively reduced decay rate, so that the system is effectively promoted to the strong coupling regime.
\end{abstract}

\maketitle

Normal--mode splitting is the hallmark of strongly coupled systems. In this regime two interacting systems exchange excitations faster than they are dissipated, and form collective normal modes
the hybridized excitations of which are superpositions of the constituent systems' excitations~\cite{Kimble,Dobrindt2008}. This regime is necessary for the observation of coherent quantum dynamics of the interacting systems and is a central achievement in research aimed at the control and manipulation of quantum systems~\cite{nms}.
In cavity opto/electro--mechanics, where electromagnetic fields and mechanical resonators interact via radiation pressure, normal--mode splitting  and strong coupling have already been obtained, using sufficiently strong power of the input driving electromagnetic field~\cite{Groeblacher-SC-2009},
or working at cryogenic temperatures with relatively large single-photon coupling~\cite{Teufel-SC-2011,Verhagen2012}.

In this letter we report on the oxymoron of observing normal--mode splitting in a weakly coupled system.
Specifically, we have designed and implemented a feedback system~\cite{Rossi2017,Kralj2017} which permits the formation of hybridized normal modes also  at room temperature and in a relatively modest device, in terms of single-photon optomechanical interaction strength (as compared to the devices used in Refs.~\cite{Groeblacher-SC-2009,Teufel-SC-2011,Verhagen2012}). Our system is basically weakly coupled
at the driving power that we can use (limited by the onset of optomechanical bistability at stronger power), and the emergence of hybridized optomechanical modes is observed when the light amplitude at the cavity output is detected and used to modulate the amplitude of the input field driving the cavity itself.
The feedback works in the anti-squashing regime, close to the feedback instability, where light fluctuations are enhanced over a narrow frequency range around the cavity resonance. 
In this regime the system behaves effectively as an equivalent optomechanical system with reduced cavity linewidth. This allows coherent energy oscillations between light and vibrational degrees of freedom when, for example, a coherent light pulse is injected into the cavity mode, similar to what has been
discussed in Ref.~\cite{Verhagen2012}. 

Light (anti--) squashing~\cite{Shapiro1987Theory-of-light,Wiseman1998In-Loop-Squeezi,Wiseman1999Squashed-states} refers to an in--loop (enhancement) reduction of light fluctuations within a (positive) negative feedback loop. 
Even if the sub-shot noise features of in-loop light disappear out of the loop, so that squashing is different from real squeezing~\cite{Shapiro1987Theory-of-light}, useful applications of in-loop light have been proposed~\cite{Wiseman1998In-Loop-Squeezi,Wiseman1999Squashed-states} and realized~\cite{Rossi2017,Kralj2017}.
In this context,
the results presented here demonstrate the potentiality of the in-loop cavity as a novel powerful tool for manipulating mechanical systems.
It can be useful in situations which require a reduced cavity decay or when, due to technical limitations, increasing the pump power is not a viable option, e.g. in case of optomechanical bistability (as in our system) or large absorption (which may lead to detrimental thermorefractive effects, in turn detuning the cavity mode~\cite{Favero2015}). Our results apply directly to the high-temperature classical regime. However, as already discussed in the case of ground state cooling~\cite{Rossi2017}, this technique can also be successfully applied to the control of mechanical resonators
at the quantum level.

Our system, described in more detail in Refs.~\cite{Rossi2017,Kralj2017,Serra2016Microfabricatio}, consists of a double--sided, symmetric, optical Fabry--P\'{e}rot cavity and a low--absorption~\cite{Serra2016Microfabricatio} circular SiN membrane in a membrane--in--the--middle setup~\cite{Thompson:2008uq}.  We focus on the fundamental mechanical mode, with resonance frequency $\wm=2\pi\times\SI{343.13}{\kilo\hertz}$ and a decay rate $\gm=2\pi\times\SI{1.18}{\hertz}$~\cite{Rossi2017,Kralj2017}.
The cavity has an empty--cavity finesse of $\mathcal{F}_0 = 42000$, corresponding to an amplitude decay rate $\kappa$
$= 2\pi \times \SI{20}{\kilo\hertz}$~\cite{Rossi2017,Kralj2017}.
Experimentally, these values are determined  by placing the membrane at a node (or an anti--node) of the cavity standing wave, since the finesse is generally diminished by the membrane optical absorption and surface roughness, and is a periodic function of its position~\cite{Serra2016Microfabricatio,Biancofiore2011}.

\begin{figure}[t]
\includegraphics[width=0.475\textwidth]{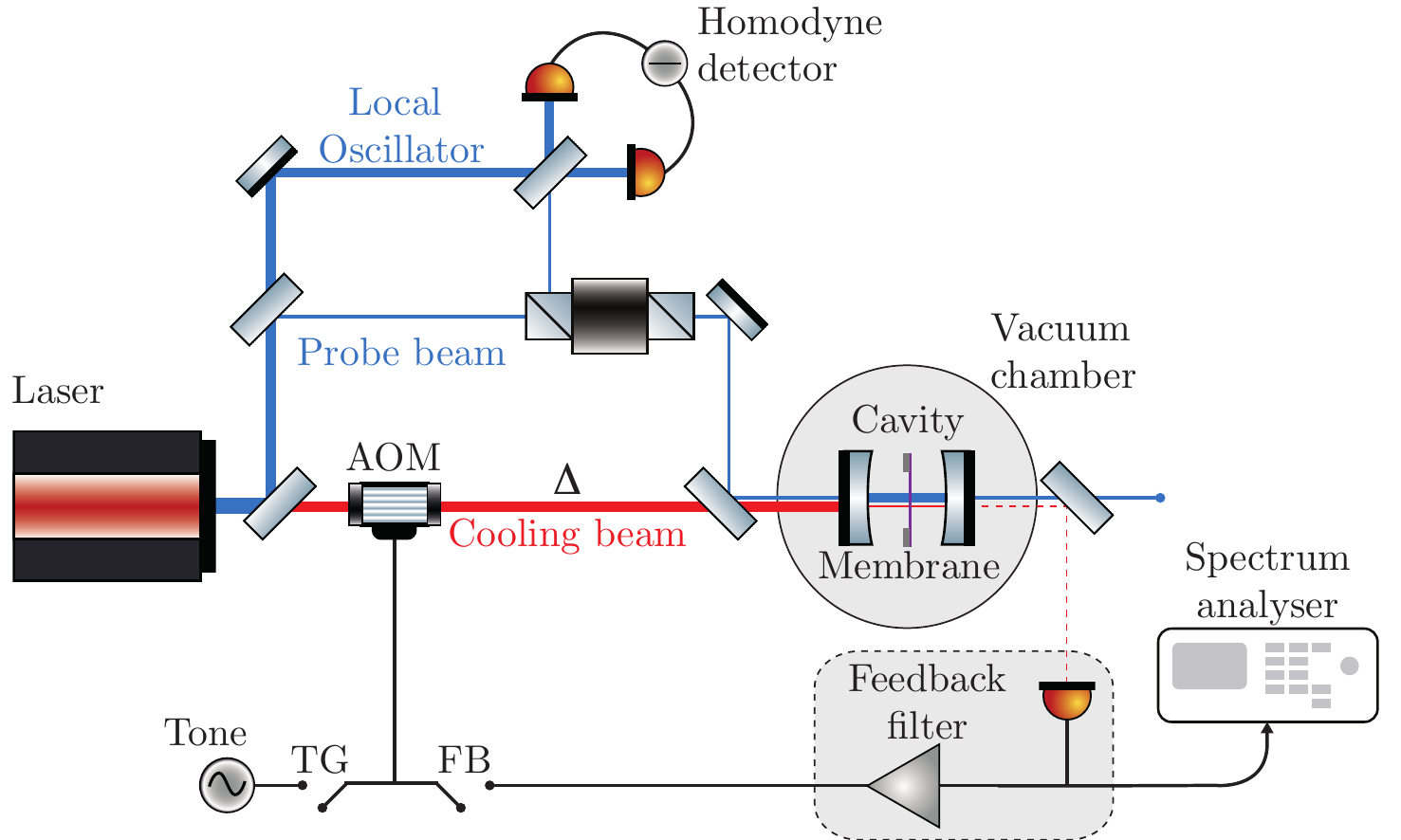}
\caption{(Color online) A \SI{1064}{\nano\meter} laser generates two beams. The probe beam, indicated by blue lines, is used to lock the laser frequency to the cavity resonance. Its phase, in which the membrane mechanical motion is encoded, is monitored with a homodyne scheme. The cooling beam, represented by red lines, provides the optomechanical interaction and is enclosed within a feedback loop. After being transmitted by the cavity, its amplitude is detected and the resulting signal is electronically processed and used to modulate the amplitude of the input field. In this way both the noise properties of light and the cavity susceptibility are modified.}
\label{fig:scheme_only_cav}
\end{figure}

The experimental setup is shown in \figurename~\ref{fig:scheme_only_cav}. Two laser
beams are utilized. The probe beam is used both to lock the laser frequency to the cavity resonance  and to monitor the cavity phase fluctuations via balanced homodyne detection.
The cooling (pump) beam, detuned by a frequency $\Delta$ from the cavity resonance by means of two
acousto--optic modulators (shown schematically as AOM in \figurename~\ref{fig:scheme_only_cav}), drives the optical cavity and provides the optomechanical interaction. This field is not a coherent, free field, but is subjected to a feedback, i.e. it is an in--loop field. After being filtered by the cavity, the amplitude quadrature of the transmitted field is directly detected with a single photodiode. The resulting photocurrent is amplified, filtered and fed back to the AOM driver in order to modulate the amplitude of the input field, thus closing the loop.
The full characterisation of the feedback response function is reported in Refs.~\cite{Rossi2017,Kralj2017}, where we have already demonstrated that this kind of feedback can be employed to enhance the efficiency of optomechanical sideband cooling. In particular we have showed how the in-loop spectra change when the feedback goes from positive to negative.

Enclosing the optical cavity within the loop~\cite{Taubman1995Intensity-feedb,Wiseman1999Squashed-states,Shapiro1987Theory-of-light,Wiseman1998In-Loop-Squeezi,Rossi2017,Kralj2017} effectively modifies its susceptibility for the in--loop optical field, such that (see also~\cite{SM})
\begin{equation}
	\tcce(\omega) = \frac{\tcc(\omega)}
		{1 - \tcfb(\omega)\left[ \tcc(\omega)\,\e^{-\im \theta_\Delta}
				+ \tccs(-\omega)\,\e^{\im \theta_\Delta}\right]}\,,
\end{equation}
where $\tcc(\omega) = [\kappa+\im(\omega-\Delta)]^{-1}$ is the cavity susceptibility, $\tcfb(\omega) = \eta\sqrt{2\ko}2\kp\sqrt{n_\mathrm{s}}\tgfb(\omega)$, with $\eta$ the detection efficiency, $\ko$ and $\kp$ the input and output cavity decay rate respectively, $n_\mathrm{s}$ the mean intracavity photon number, and $\tgfb(\omega)$ the feedback control function [$\tgfb^\ast(-\omega)= \tgfb(\omega)$].
Furthermore, the dimensionless displacement of the mechanical oscillator measured by the out--of--loop probe beam, $\delta\tilde{q}= \tilde\chi_\mathrm{m}^\mathrm{o,eff}(\omega)[\tilde{\xi}(\omega)  + \wt{\mathcal{N}}^\mathrm{eff}(\omega) ]$~\cite{SM}, is the sum of a term proportional to thermal noise, described by the zero mean stochastic noise operator $\tilde\xi(\omega)$, and a term due to the interaction with the cavity, proportional to radiation pressure noise, reshaped by the effective cavity susceptibility according to the relation
$\wt{\mathcal{N}}^\mathrm{eff}(\omega) = G \{ \tcce(\omega)\,\tilde n + [\tcce(\omega)]^\ast\,\tilde n^\dag \}$,
with $\tilde{n}$ the radiation pressure noise operator~\cite{SM} and $G =g_0\sqrt{2 n_\mathrm{s}}$ the (many--photon) optomechanical coupling strength~\cite{Dobrindt2008,Genes2008}, where $g_0$ is the single--photon optomechanical coupling.
Finally, in the expression for the mechanical displacement, the factor $\tilde\chi_\mathrm{m}^\mathrm{o,eff}(\omega)$ is the modified mechanical susceptibility that is dressed by the
effective self--energy $\Sigma^\mathrm{eff}(\omega)= -\im G^2 \{ \tcce(\omega) - [\tcce(-\omega)]^\ast\}$ according to
\begin{align}
	[\tilde\chi_\mathrm{m}^\mathrm{o,eff}(\omega)]^{-1} =
		[\tcm(\omega)]^{-1} + \Sigma^\mathrm{eff}(\omega)\,,
\end{align}
where the bare susceptibility is
$[\tcm(\omega)]^{-1} = (\wm^2 - \omega^2 -\im \omega \gm)/\wm$.

In the resolved sideband limit, $ \omega_{\rm{m}} \gg \kappa$,
and for $\Delta \sim \omega_{\rm{m}}$ in order to cool the resonator, the effective cavity susceptibility for
frequencies close to the cavity resonance $\omega \sim \Delta$ can be approximated as
$\tcce(\omega) \sim [\kappa_{\mathrm{eff}} + \im( \Delta_{\mathrm{eff}} - \omega)]^{-1}$,
where $\kappa_{\mathrm{eff}}=\kappa + \Im [\tcfb(\Delta)]$ and $\Delta_{\mathrm{eff}}=\Delta-\Re[\tcfb(\Delta)]$.
These relations allow to significantly simplify the expressions reported above and interpret the system dynamics in terms of that of a standard optomechanical system with a modified cavity.
In particular, in the positive feedback regime
(corresponding to light anti--squashing) the in--loop optical mode experiences an effectively reduced decay rate, which tends to zero as the feedback gain is increased and approaches the feedback instability~\cite{Rossi2017,Kralj2017}. This in turn amounts to an increased optomechanical cooperativity $C_\eff = 2\, G^2/\kappa_\eff \, \gm$.  In Refs.~\cite{Rossi2017,Kralj2017} we have correspondingly shown that this effect can be employed to augment the mechanical damping rate  $\Gamma_\eff$ and hence to improve sideband cooling of mechanical motion.
Here we demonstrate that in--loop optical cavities represent a new, powerful tool for reaching the strong coupling regime, owing to an effective reduction of the cavity linewidth $\kappa_\eff$.

Normal--mode splitting
is a clear signature of strong coupling, being that it is only observable above the threshold $G \gtrsim \kappa_\eff$~\cite{Dobrindt2008,Groeblacher-SC-2009} (in typical optomechanical systems the other condition $G > \gamma_\mathrm{m}$ is easily satisfied).
Since both normal modes are combinations of light and mechanical modes, they are both visible in the
detectable mechanical displacement spectrum as distinct peaks at frequencies  $\omega_\pm$, separated by $\omega_+ - \omega_- \simeq \sqrt{2}G$ when $\Delta_{\rm eff}=\omega_{\rm m}$. The two peaks are distinguishable if the corresponding linewidths, which are of the order of $\kappa_{\rm eff}$, are smaller than $G$.
\begin{figure}[b]
\includegraphics[width=0.475\textwidth]{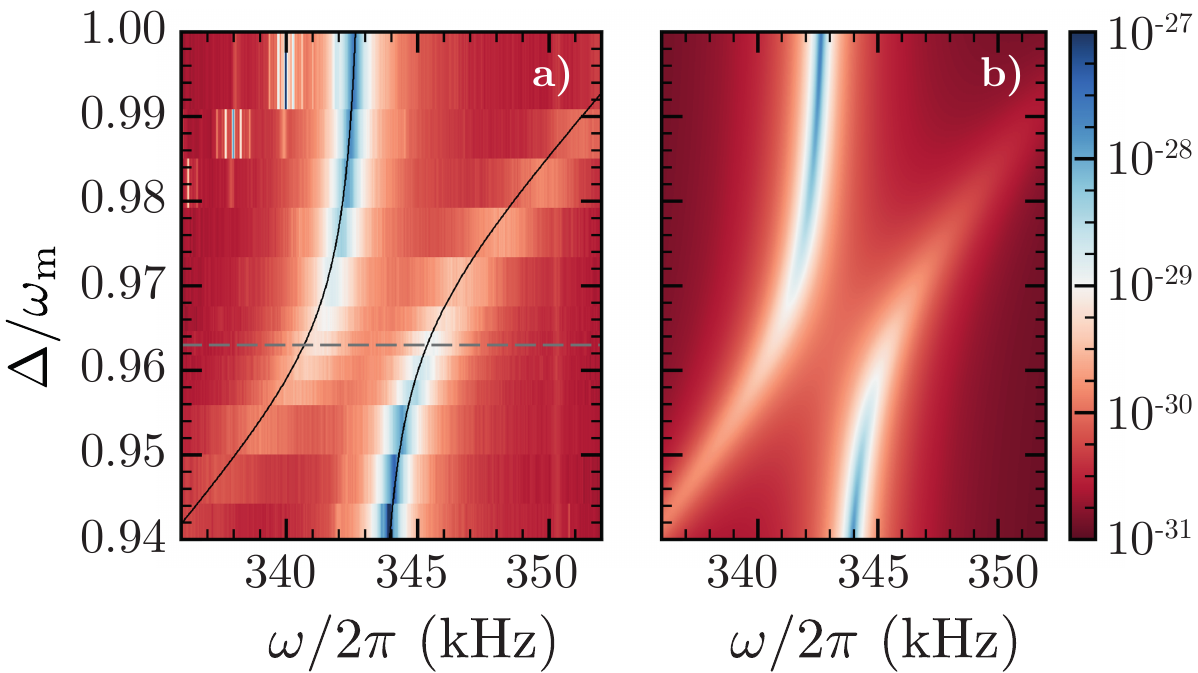}
\caption{ (Color online) Normal mode splitting. a), Measured, and b), theoretically predicted splitting of the fundamental mechanical mode in the strong-coupling regime as a function of detuning, with the two normal modes exhibiting avoided crossing. The dashed grey line indicates the optimal value of the detuning for sideband--cooling with feedback. The values of the colour scale are in m$^2$/Hz and correspond to the displacement spectral noise evaluated as $S_{xx}(\omega) = x_0^2\,S_{qq}(\omega)$ with $x_0 = \sqrt{\hbar/2m\wm}$ the zero point motion factor, and $S_{qq}(\omega)$ the power spectrum of the dimensionless displacement operator $\delta q$~\cite{SM}.
}
\label{fig:normal_mode_splitting}
\end{figure}
In particular, strong coupling manifests itself as avoided crossing for the values of the normal frequencies $\omega_\pm$ when the cavity detuning is varied. This is apparent from \figurename~\ref{fig:normal_mode_splitting}, showing the spectra of the displacement fluctuations of the mechanical mode interacting with the in--loop optical mode, recorded via homodyne detection of the probe beam. In \figurename~\ref{fig:normal_mode_splitting}a) a color--plot is used to show these spectra as a function of frequency and normalised detuning, acquired with the maximum attainable feedback gain, and panel b) is the theoretical expectation. The parameters used for the simulation, determined independently, are the decay rate $\kappa = 2\pi \times \SI{22}{\kilo\hertz}$,
the single--photon optomechanical coupling estimated to be $g_0 = 2\pi\times\SI{1.8}{\hertz}$ at this membrane position, and the input cooling power $P=\SI{10}{\micro\watt}$.
These parameters correspond to $G \sim 2 \pi \times 3836$ Hz, which is larger than $\gamma_{\rm m}$,  but lower
than $\kappa$, implying that the optomechanical system is initially far from the
strong--coupling regime. The feedback is then set to operate in the anti--squashing regime, with such a value of gain that the threshold $G \sim \kappa_\eff$ is surpassed and normal mode splitting becomes visible.
\begin{figure}[b]
\includegraphics[width=0.4\textwidth]{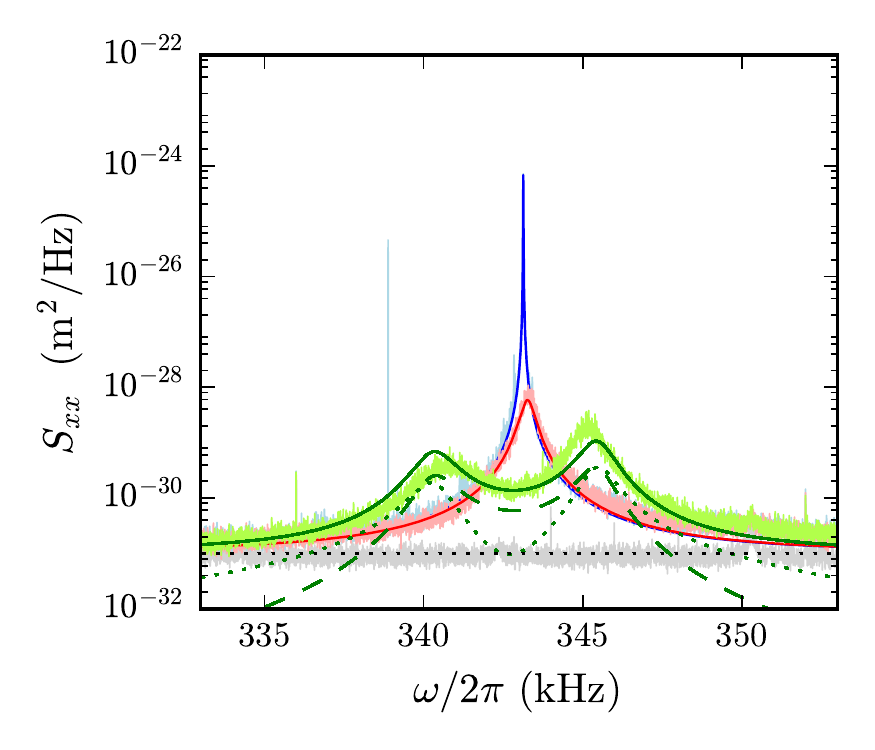}
\caption{(Color online) Displacement spectral noise, $S_{xx}(\omega) = x_0^2\,S_{qq}(\omega)$ with $x_0 = \sqrt{\hbar/2m\wm}$ and offset by the shot--noise grey trace,  of the (0,1) membrane mode at room temperature (blue trace), and sideband--cooled (red trace) with a pump of $P = \SI{10}{\micro\watt}$ detuned by $\Delta = \SI{330}{\kilo\hertz}$. Increasing the gain with the feedback operating in the anti--squashing regime effectively reduces the cavity linewidth, allowing to enter the strong--coupling regime, as seen from the appearance of two hybrid modes (green trace).
The green--solid line represents the theoretical expectation according to eq.~\eqref{eq:Sqq_aprox}, and is the sum of
the comparable thermal and feedback terms shown as dotted-- and dashed--line, respectively, while the radiation pressure contribution is negligible. The narrow feature at $\sim\SI{339}{\kilo\hertz}$ is a calibration tone.}
\label{fig:spectrum}
\end{figure}

Let us now analyse these spectra in more detail.
In the resolved sideband limit, the symmetrised displacement noise spectrum can be expressed as~\cite{SM}
\begin{eqnarray}\label{eq:Sqq_aprox}
S_{qq}(\omega)\simeq\abs{\tcmoe(\omega)}^2\ [S_{\rm th}+S_{\rm rp}^{\rm eff}(\omega)+S_{\rm fb}(\omega)]\,,
\end{eqnarray}
where the first two terms account for the standard spectrum (with no feedback) for an optomechanical system, but with cavity decay rate $\kappa_{\rm eff}$, and the last term can be interpreted as additional noise due to the feedback and is given by~\cite{SM}
\begin{eqnarray}
	S_{\rm fb}(\omega)\sim
		 G^2\ \ZZ^\Delta\,[ \abs{\tcce(\omega)}^2+\abs{[\tcce(-\omega)]}^2] \ ,
\end{eqnarray}
which has the same form of the radiation pressure term, except for the factor $\ZZ^\Delta = [(\Delta-\Delta_{\rm eff})^2+(\kappa_{\rm eff}-\kappa)^2]/2\eta\kappa'$ replacing $\kappa_{\rm eff}$.
\figurename~\ref{fig:spectrum} shows the spectrum of the fundamental mechanical mode excited by thermal fluctuations at \SI{300}{\kelvin} (blue trace), with an optomechanical contribution due to the quasi--resonant probe beam with $\SI{15}{\micro\watt}$ of power, which slightly cools down the mechanical mode, increasing the damping rate by a factor of $\sim2.8$, due to an estimated probe detuning of around $2\pi\,\times\,$\SI{300}{\hertz}.
The red trace demonstrates the standard (no feedback) sideband--cooling due to the cooling beam with a detuning set to $\Delta = 2\pi\times\SI{330}{\kilo\hertz}$, and the other optomechanical parameters set as for the data in \figurename~\ref{fig:normal_mode_splitting}, such that the strong coupling regime is initially not reached.
Finally, the green trace corresponds to the cross--section of \figurename~\ref{fig:normal_mode_splitting}a) indicated by the grey dashed line.
In this particular case we estimate, from the experimental data and the simulation, the effective parameters
$\kappa_\mathrm{eff}\sim 2\pi\times\SI{1210}{\hertz}$ and $\Delta_\mathrm{eff}\sim 2\pi\times\SI{342.65}{\kilo\hertz}$.
Since $\ZZ^\Delta \gg \kappa_{\rm eff}$ in the range of parameters relevant to our experiment, the feedback noise, differently from the radiation pressure term, provides a non-negligible contribution to the overall spectrum with respect to the thermal one, as indicated by the dashed and dotted lines.

\begin{figure}[t]
\includegraphics[width=0.4\textwidth]{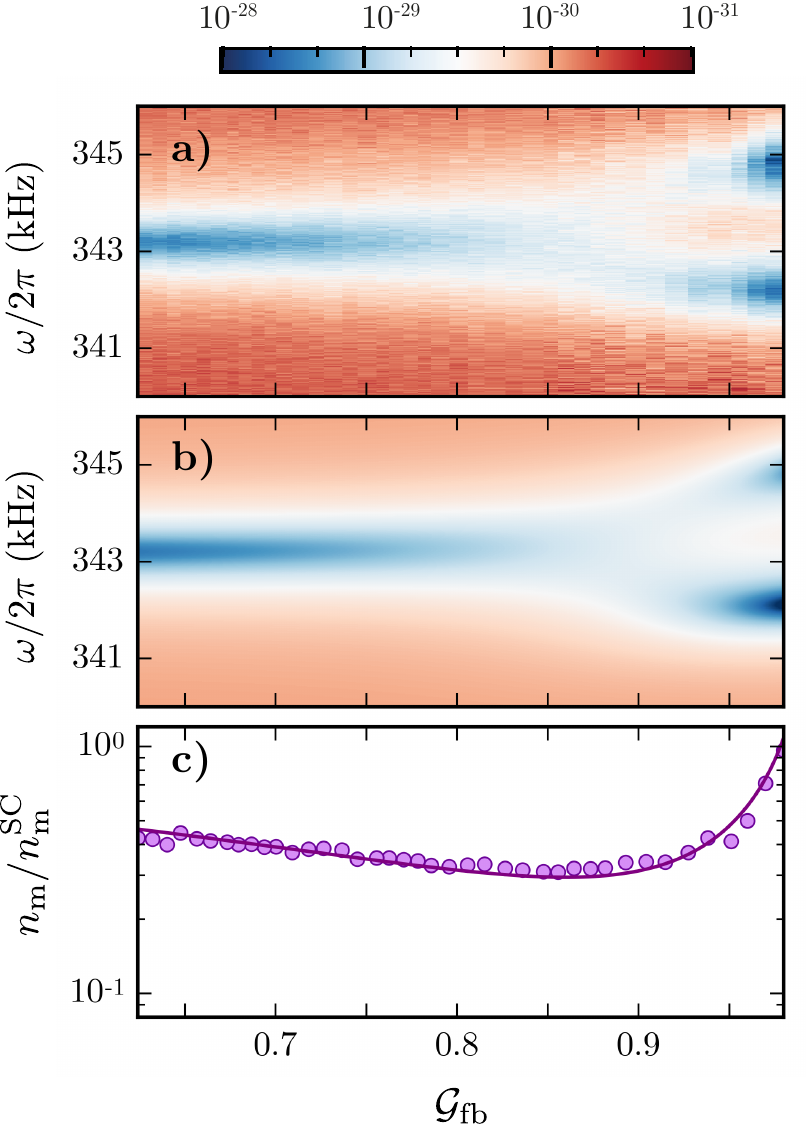}
\caption{(Color online) Transition between the weak-- and strong--coupling regime. a), Power spectra of mechanical displacement fluctuations varying the feedback gain $\mcg_\fb$, and b), the corresponding simulation
evaluated as $S_{xx}(\omega) = x_0^2\,S_{qq}(\omega)$ with $x_0 = \sqrt{\hbar/2m\wm}$. The color scale is shown at the top in \si[per-mode=symbol]{\meter\squared\per\hertz}. At low gain the mechanical motion is described by a single mode the dynamics of which is modified by the in--loop optomechanical interaction. At high gain, instead, the spectrum becomes double--peaked: the strong interaction produces hybridized optomechanical modes and the mechanical motion is a superposition of these two normal modes.
The difference in frequency between the two peaks in the spectrum with maximum gain corresponds to $G = 2\pi\times \SI{1.87}{\kilo\hertz}$. The parameters for this measurement ($P$, $\Delta$, $\kappa$ and $g_0$) yield $G = 2\pi \times \SI{1.96}{\kilo\hertz}$, in good agreement with the experimental estimation.
c), The ratio of the effective phonon number with and without feedback. The circles are obtained by numerical integration of the measured spectra, while the solid line corresponds to Eq.~\rp{eq:nm} evaluated using the measured parameters, which is valid both in the weak-- and in the strong--coupling regime. The feedback scheme enhances the cooling rate with respect to standard sideband--cooling by a factor of \SI{5}{\decibel}. }
\label{fig:strong_coupling}
\end{figure}

The results we have presented are obtained in a condition in which the pump field efficiently cools the mechanical resonator~\cite{Rossi2017,Kralj2017}. In general, when an optomechanical system enters the strong coupling regime, the efficiency of sideband cooling decreases.
Hereafter we report on the similar effect that we observe as we increase
the feedback gain towards instability, while keeping the other parameters fixed, as shown in~\figurename~\ref{fig:strong_coupling}. Panel~a) presents a plot of the mechanical displacement spectra as a function of frequency and feedback gain $\mcg_\fb =-\Im [\tcfb(\Delta)]/\kappa$, normalised in such a way that $\mcg_\fb = 1$ when $\kappa_\eff = 0$, i.e. at the feedback stability threshold. In panel~b) we report the corresponding and consistent results simulated using the theoretical model with the previously listed parameters for the membrane mode, $P = \SI{26}{\micro\watt}$ and $\Delta = 2\pi \times \SI{334.9}{\kilo\hertz}$ for the optical pump, $\kappa = 2 \pi \times \SI{21}{\kilo\hertz}$ and $g_0 = 2\pi \times \SI{0.6}{\hertz}$.
Finally, panel c) shows that at low gain values the cooling efficiency increases with the feedback gain. As explained previously, this effect can be understood as a result of
the increment in the optomechanical cooperativity due to the effectively reduced
in--loop cavity decay rate. We further note that, as expected, the enhanced cooperativity does not imply an improvement of optical cooling all the way towards the instability point. Rather, the cooling works well in the weak--coupling limit, i.e. when the cavity response time $\kappa_\eff^{-1}$ is shorter than the decay time of the oscillator modified by the optomechanical interaction $\Gamma_\eff^{-1}$, so as to allow mechanical thermal energy to be transferred into the cavity mode and leak out~\cite{Genes2008}. Conversely, around the threshold $\kappa_\eff \sim G$ the overall mechanical damping rate is of the order of the cavity linewidth, $\Gamma_\eff \sim \kappa_{\rm eff}$, and as
the gain is increased further, $\Gamma_\eff$ grows, while $\kappa_\eff$ gets smaller, such that the cooling efficiency decreases.
In particular, for high temperature, in the resolved sideband limit, $\kappa_{\rm eff}\ll\Delta_{\rm eff}\sim\omega_{\rm m}$, small optomechanical coupling $G\ll\omega_{\rm m}$, and a small mechanical decay rate  $\gamma_{\rm m}\ll (\Gamma_{\rm eff},\kappa_{\rm eff})$, the steady state average number of mechanical excitations $n_{\rm m}$ can be evaluated in terms of the integral of the spectrum $S_{qq}(\omega)$~\cite{Genes2008}, and it is given by~\cite{SM}
\begin{eqnarray}
	n_{\rm m}\sim
	n_{\rm m}^{\rm th,eff}\ \frac{\gamma_{\rm m}}{\Gamma_{\rm eff}}\pt{1+ \frac{\Gamma_{\rm eff}}{2\,\kappa_{\rm eff}}}\,,
	\label{eq:nm}
\end{eqnarray}
which is equal to the result for a standard optomechanical system (with no feedback), but with cavity decay rate $\kappa_{\rm eff}$, and  in a higher temperature reservoir $n_{\rm m}^{\rm th,eff} \sim n_{\rm m}^{\rm th}+ n_{\rm m}^{\rm eff}$\,, with
\begin{eqnarray}
	n_{\rm m}^{\rm eff} \sim  \frac{\ZZ^\Delta\ \Gamma_{\rm eff}}{\gamma_{\rm m}\pt{2\,\kappa_{\rm eff}+\Gamma_{\rm eff}}}\ .
\end{eqnarray}
The validity of this result is  demonstrated in \figurename~\ref{fig:strong_coupling}c) where we report the effective phonon number of the mechanical mode, normalised with respect to the occupancy obtained by standard sideband--cooling without feedback, $n_\mathrm{m}^\mathrm{SC}$.
In particular, the solid line, which is in very good agreement with the data (dots), represents the expected average phonon number defined in  Eq.~\eqref{eq:nm}.  The optimal cooling gain is $\mcg_\fb \approx 0.9$, and beyond this value the spectrum becomes double--peaked [\figurename~\ref{fig:strong_coupling}a), and b)], indicating that the system enters the strong coupling regime.

To conclude, we emphasise that, as demonstrated by our results, feedback--controlled light represents a promising approach to the control of the optomechanical dynamics which offers the possibility to tune the effective cavity linewidth at will.
In particular, herein we have shown that this allows to access the regime of strong coupling, characterised by the emergence of hybridized normal modes, even when the optomechanical interaction is small as compared to the natural dissipation rates, so that the original system is in fact weakly coupled.
In our experiment, using the optimal parameters of
\figurename~\ref{fig:normal_mode_splitting}, the effective cavity decay rate is reduced by a factor $20$, and the system is promoted to the strong coupling regime with an estimated cooperativity parameter of $C_{\rm eff}\simeq 2\times10^4$.
We further note that the ability to effectively reduce the cavity linewidth may ease tasks such as transduction, storage and retrieval  of signals and energy~\cite{Barzanjeh2012,Bagci2013,Reed2017}
with low frequency  massive resonators.
Finally, this technique could also be exploited to improve certain protocols for the preparation of non-classical mechanical states~\cite{entanglement}, which are more efficient at low cavity decay rate, or to enhance the efficiency of mechanical heat engines which work in the strong coupling regime~\cite{Zhang} or which make use of correlated reservoirs~\cite{Klaers}.

{\it Acknowledgments -- }
We acknowledge the support of the European Commission through the H2020-FETPROACT-2016 project~n.~732894~``HOT''.

\clearpage
\begin{widetext}
\begin{center}
\textbf{\large Supplemental Material: \\Normal--mode splitting in a weakly coupled optomechanical system}

\vspace{5mm}
{\normalsize Massimiliano Rossi,\textsuperscript{1, 2, \textcolor{blue}{*}} Nenad Kralj,\textsuperscript{2} Stefano Zippilli,\textsuperscript{2, 3} Riccardo Natali,\textsuperscript{2, 3} Antonio Borrielli,\textsuperscript{4}

Gregory Padraud,\textsuperscript{5} Enrico Serra,\textsuperscript{5, 6} Giovanni Di Giuseppe,\textsuperscript{2, 3,\textcolor{blue}{$\dagger$}} and David Vitali\textsuperscript{2, 3, 7\textcolor{blue}{$\ddagger$}}}

\vspace{2mm}
\textsuperscript{1}\textit{School of Higher Studies ``C. Urbani", University of Camerino, 62032 Camerino (MC), Italy}

\textsuperscript{2}\textit{School of Science and Technology, Physics Division, University of Camerino, 62032 Camerino (MC), Italy}

\textsuperscript{3}\textit{INFN, Sezione di Perugia, 06123 Perugia (PG), Italy}

\textsuperscript{4}\textit{Institute of Materials for Electronics and Magnetism,}

\textit{Nanoscience-Trento-FBK Division, 38123 Povo (TN), Italy}

\textsuperscript{5}\textit{Delft University of Technology, Else Kooi Laboratory, 2628 Delft, The Netherlands}

\textsuperscript{6}\textit{Istituto Nazionale di Fisica Nucleare, TIFPA, 38123 Povo (TN), Italy}

\textsuperscript{7}\textit{CNR-INO, L.go Enrico Fermi 6, I-50125 Firenze, Italy}
\end{center}
\end{widetext}
\setcounter{equation}{0}
\setcounter{figure}{0}
\setcounter{table}{0}
\setcounter{page}{1}
\makeatletter
\renewcommand{\theequation}{S\arabic{equation}}
\renewcommand{\thefigure}{S\arabic{figure}}
\renewcommand{\bibnumfmt}[1]{[S#1]}
\renewcommand{\citenumfont}[1]{S#1}

\section{Theory}
\begin{figure}[th!]
\includegraphics[width=0.375\textwidth]{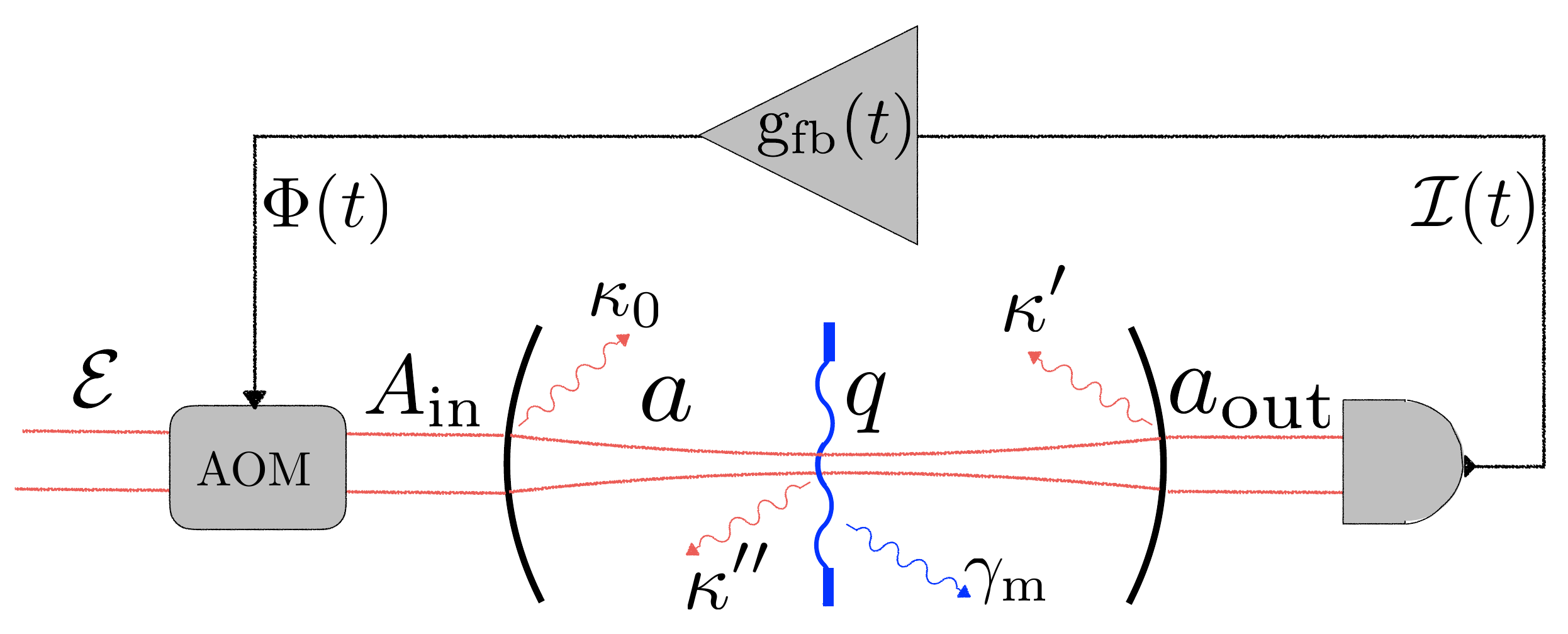}
\caption{Sketch of the optomechanical system.}
\label{fig:Sketch}
\end{figure}

We consider a mode of a mechanical resonator described by the dimensionless position and momentum operators $q$ and $p$ (with $\pq{q,p}=\ii$) with  frequency $\omega_{\rm m}$, decay rate $\gamma_{\rm m}$, and mass $m$. 
The operators $q$ and $p$ are related to the real position $\hat x=x_0\,q$ and momentum $\hat P=p_0\,p$, by the factors $x_0=\sqrt{\hbar/2m\omega_\mathrm{m}}$ and $p_0=\sqrt{\hbar m\omega_\mathrm{m}/2}$. The mechanical resonator is coupled to a resonant mode of a Fabry--P\'{e}rot cavity, with operators $a$ and $a\da$ ($[a,a\da]=1$), at frequency $\omega_c$ and with decay rate $\kappa$, via radiation pressure with strength $g_0 = -x_0d\omega_c/dx$. The cavity is driven by a laser field at frequency $\omega_L$, amplitude modulated by a feedback system, which measures the light transmitted by the cavity (see \figurename~\ref{fig:Sketch}).
The equations of motion for this system  are
\begin{align}
  \dfq &=\wm p	\label{eq:dq}\,, \\
  \dfp &= -\wm q - \gm p  + \sqrt{2}g_0 a^\dag a + \xi \label{eq:dp}\,,\\
  \dfa &= -(\kappa + \im \Delta_0)\,a + \im \sqrt{2}g_0 a q + \sqrt{2\ko}\,A_{\rm in}\,\e^{-\im \theta_\Delta} +\nonumber\\
  						 &\hspace{3.65cm}+ \sqrt{2\kp}\,\ainp
  						 + \sqrt{2\kpp}\,\ainpp\,,   \label{eq:da}
\end{align}
where $\Delta_0 = \wc-\omega_L$ is the beam detuning, and the phase $\theta_\Delta = \arctan(-\Delta/\kappa)$ accounts for having chosen the phase of the cavity field as reference, with $\Delta$ the effective detuning defined below. The operator $\xi$ describes thermal noise acting on the mechanical resonator and, in the high temperature limit relevant here, is characterized by the correlation function $\av{\xi(t)\ \xi(t')}=\gm\pt{2\,n_{\rm th}+1}\delta(t-t')$ \cite{Giovannetti2001sm}, with $n_{\rm th}$ the number of thermal excitations.
Moreover, we decompose the total cavity decay rate as $\kappa = \ko + \kp + \kpp$, in terms of the contributions due to
the losses of the input mirror $\ko$, output mirror $\kp$ and additional internal losses $\kpp$.
Correspondingly we have introduced three input operators $A_{\rm in}$, $\ainp$ and $\ainpp$. In particular, the latter two describe vacuum noise and are characterized by the correlation function $\av{\ainp(t)\ {\ainp}\da(t')}=\av{\ainpp(t)\ {\ainpp}\da(t')}=\delta(t-t')$, instead the operator associated to the field at the input mirror can be decomposed as
\begin{eqnarray}
A_{\rm in}=\ain+\EE+\Phi\,,
\end{eqnarray}
where $\ain$ describes vacuum noise, $\mathcal{E} = \sqrt{\mathcal{P}/\hbar\omega_L}$ accounts for the pump field at power $\PP$, and $\Phi$ is the contribution due to the feedback.

This last part can be expressed as
\begin{equation}
	\Phi(t)= \int_{-\infty}^t \dd t^\prime \, \gfb(t-t^\prime)\, \mathcal{I}(t^\prime)\,,
\end{equation}
where $\gfb(t)$ is the causal filter function of the feedback and $\II(t)$ describes the detection photocurrent
given by
\begin{align}
    \mathcal{I}(t)&= \left[\sqrt{\eta}\,{a^{\dagger}_{\rm out}} +
    			\sqrt{1-\eta} c^\dag\right] \left[\sqrt{\eta}\,{a_{\rm out}} + \sqrt{1-\eta} c\right] \nonumber\\
			&=\eta\,{a^{ \dagger}_{\rm out}} {a_{\rm out}}
			+ \sqrt{\eta(1-\eta)}\,({a^{\dagger}_{\rm out}} c + c^\dag {a_{\rm out}})+ (1-\eta)\,c^\dag c\,,
\end{align}
where $\eta$ is the detection efficiency, $c$ represents additional vacuum noise due to the inefficiency of the detection, and ${a_{\rm out}}$  describes the transmitted output field given by
\begin{equation}
	{a_{\rm out}} = \sqrt{2\kappa^\prime}\,a - \ainp\,.
\end{equation}
Approximate solutions of Eqs.~\eqref{eq:dq}--\eqref{eq:da}
can be found, provided the system is stable, by linearisation of the system for
small fluctuations, $\delta a$ and $\delta q$, around the steady state solution $\alpha_s=a-\delta a$ and $q_s=q-\delta q$. These values are determined by imposing $\av{\dot{q}}  = \av{\dot{a}} = 0$, and are given by
\begin{align}
	q_\mathrm{s} = \sqrt{2}g_0\frac{\as^2}{\wm}
		\hspace{1cm}
	\as = \frac{\sqrt{2\kappa_0}}{|\kappa +\im\Delta|}(\mathcal{E} + \bar{\Phi})\,,
\end{align}
where we have
introduced the effective detuning
$\Delta = \Delta_0 - \sqrt{2}g_0\,q_\mathrm{s} = \Delta_0 -  2g_0^2\, \as^2/\wm$, and the averaged feedback response
$\bar{\Phi}=2\eta\kappa^\prime\, \as^2\,\int_0^{\infty} \dd t \gfb(t)$.

The linearised equations for the fluctuations (obtained by neglecting contributions at second order in the fluctuations) are
\begin{align}
    \delta \ddot q&= -\wm^2\, \dtq - \gm\, \delta\dot q  + \wm\sqrt{2}g_0\,\as\,( \dta +  \dta^\dag ) + \wm\,\xi\,, 	\\
    \delta \dot{a} &= -(k + \im \Delta)\,\dta + \im\, \sqrt{2}g_0\, \as \,\dtq
  		+ \sqrt{2\kappa_0}\,\delta \Phi_{\rm a}\,\e^{-\im\theta_\Delta} +
				\nonumber \\ & \hspace{2.7cm}
		+ \sqrt{2\kappa_0}\,(\delta \Phi_{\rm n} + \ain)\,\e^{-\im\theta_\Delta}+
				\nonumber \\ & \hspace{2.75cm}
  		+ \sqrt{2\kp}\,\ainp  + \sqrt{2\kpp}\,\ainpp \,,
\end{align}
where we have eliminated the equation for $p$, and the feedback response $\Phi$ has been decomposed as $ \Phi = \bar{\Phi} +  \delta \Phi_{\rm a} +  \delta \Phi_{\rm n}$ with
\begin{align}
    \delta \Phi_{\rm a} &=
    		2\eta\kp\,\as \int_{-\infty}^t \dd t^\prime \gfb(t-t^\prime)\left(\dta(t^\prime) + \dta^\dag (t^\prime)\right)\,, \\
    \delta \Phi_{\rm n} &=
    		-\eta\,\sqrt{2\kp}\,\as \int_{-\infty}^t \dd t^\prime \gfb(t-t^\prime)\left(\ainp(t^\prime)
						+ a_\mathrm{in}^{\prime\,\dag}(t^\prime) \right) +
							\nonumber\\&\hspace{.5cm}
						+ \sqrt{\eta(1-\eta)}\sqrt{2\kp}\,\as\int_{-\infty}^t \dd t^\prime \gfb(t-t^\prime)\left(c (t^\prime)
						+ c^\dag(t^\prime) \right)\,.
\end{align}
In the frequency domain, where operators are indicated by the tilde symbol $\ \tilde{}\ $,
we find 
\begin{align}\label{qleom}
  -\im\omega\,\delta \tilde{a} &= -(k + \im \Delta)\,\delta \tilde{a} + \im\, G \,\delta \tilde{q}
  		+ \sqrt{2\ko}\,\delta \tilde{\Phi}_{\rm a}\,\e^{-\im\theta_\Delta} + \tilde{n}\,,
  		\nn\\
  -\omega^2 \delta \tilde{q}&= -\wm^2 \delta \tilde{q} +\im\omega\, \gm\,\delta \tilde{q}
  			+ G\,\wm(\delta \tilde{a} + \delta \tilde{a}^\dag )
		+\wm\, \tilde{\xi}\,,
\end{align}
where  $G = g_0\sqrt{2 n_\mathrm{s}}$ is the optomechanical coupling with $n_\mathrm{s} = \as^2$ the mean intracavity photon number,
\begin{align}\label{noise_n}
	\tilde{n}
		 &= \sqrt{2\ko}\,(\delta  \tilde{\Phi}_{\rm n} + \tilde a_{\rm in})\e^{-\im\theta_\Delta}
  		+ \sqrt{2\kp}\,\tilde a'_{\rm in}
		+ \sqrt{2\kpp}\,\tilde a''_{\rm in}\,,
\end{align}
and
\begin{align}
    \delta  \tilde{\Phi}_{\rm a} &= \eta\,2\kp\,\as\, \tgfb(\omega)\left(\delta \tilde{a}
				+ \delta \tilde{a}^\dag \right)\,,\\
    \delta \tilde{\Phi}_{\rm n} &= -\eta\,\sqrt{2\kp}\,\as\,\tgfb(\omega)\left(\tilde{a}_\mathrm{in}^\prime
						+ \tilde{a}_\mathrm{in}^{\prime\,\dag} \right) \,+
							\nonumber\\&\hspace{.5cm}
			+ \sqrt{\eta(1-\eta)}\sqrt{2\kappa^\prime}\,\as\ \tilde{g}_{\rm fb}(\omega)\left( \tilde{c}
						+ \tilde{c}^\dag\right) 	\,,	
\label{Phi_n}
\end{align}
with $\tgfb(\omega)$ the Fourier transform of the filter function which fulfils the relation $\tgfb(\omega)^*=\tgfb(-\omega)$.
We note that when considering operators in Fourier space the symbol ${}\da$ is not used to indicate the hermitian conjugate of the corresponding operator, rather, the hermitian conjugate of the operator at the opposite frequency, so that given a generic operator $\tilde o\equiv\tilde o(\omega)$, then $\tilde o\da\equiv\pq{\tilde o(-\omega)}\da$.

\subsection{System response functions}

Let us now introduce some quantities that will be useful in the following discussion:
 the bare cavity susceptibility
\begin{equation}
	\tcc(\omega) = \left[ \kappa + \im(\Delta-\omega)\right]^{-1}\,,
\end{equation}
the bare mechanical susceptibility
\begin{equation}
	\tcm(\omega) = \wm\left[ \wm^2 - \omega^2 -\im\omega\gm\right]^{-1}\,,
\end{equation}
the effective cavity susceptibility modified by the feedback
\begin{equation}
	\tcce(\omega) =
	\frac{\tcc(\omega)}
		{1 - \tcfb(\omega)\left[ \tcc(\omega)\,\e^{-\im \theta_\Delta}
				+ \tccs(-\omega)\,\e^{\im \theta_\Delta}\right]}\,,
\end{equation}
the effective dressed mechanical susceptibility modified  by the optomechanical  interaction and by the feedback
\begin{equation}
	\tcm^{\rm o,eff}(\omega) = \pg{
	\tcm(\omega)^{-1}  + \Sigma^\mathrm{eff}(\omega)}^{-1}\,,
\end{equation}
with 
\begin{equation}
   \Sigma^\mathrm{eff}(\omega) = -\ii\,G^2\pq{\tcce(\omega)-\tcce(-\omega)^*}\,,
\end{equation}
and the rescaled filter function
\begin{equation}
	\tcfb(\omega) =\eta\ \sqrt{2\ko}\,2\kp\,\as \,\tgfb(\omega)\,.
\end{equation}
We finally note that,
in the resolved sideband limit $\kappa\ll\omega_m$ and for short feedback delay time $\kappa\ll 1/\tau_{\rm fb}$, the effective cavity susceptibility $\tcce(\omega)$ can be approximated, for frequencies around the cavity resonance $\omega\sim\Delta$, as the bare susceptibility of a cavity with effective cavity decay rate $\kappa_{\rm eff}$ and detuning $\Delta_{\rm eff}$, such that~\cite{Rossi2017sm}
\begin{align}\label{chi_ceff}
	\tcce(\omega)
	\sim
	\frac{1
	}{\kappa_{\mathrm{eff}}+\ii(\Delta_{\mathrm{eff}}-\omega)}\,.
\end{align}
The values of $\kappa_{\rm eff}$ and detuning $\Delta_{\rm eff}$ can be expressed in terms of the feedback filter function (which is slowly varying and essentially constant over the cavity linewidth)  evaluated for frequencies close to the cavity resonance $\tcfb(\omega)\Bigl|_{\omega\sim\Delta}\equiv \tcfb^\Delta$, as
$\kappa_{\mathrm{eff}}=\kappa+\Im \pq{\tcfb^\Delta}$ and $\Delta_{\mathrm{eff}}=\Delta-\Re\pq{\tcfb^\Delta}$,
that is
\begin{eqnarray}\label{chi_fb}
\tcfb^\Delta\sim\Delta-\Delta_{\rm eff}+\ii\pt{\kappa_{\rm eff}-\kappa}\ .
\end{eqnarray}

\subsection{Power spectrum of the mechanical position operator}

An expression for the displacement operator $\delta\tilde q$ can be derived solving Eq.~\rp{qleom}. It reads
\begin{align}\label{delta_q}
	\delta\tilde{q} = \tcm^{\rm o,eff}(\omega)\,\left[
		\tilde{\xi}(\omega)  + \widetilde{\mathcal{N}}^\mathrm{eff}(\omega) \right]\,,
\end{align}
where
\begin{equation}
	\widetilde{\mathcal{N}}^\mathrm{eff}(\omega) = G \left[ \tcce(\omega)\,\tilde n + [\tcce(-\omega)]^*\,\tilde n^\dag \right]\,.
\end{equation}
Using this expression it is possible to evaluate the
power spectrum of the mechanical displacement operator $\delta\tilde q(\omega)$ that can be measured by sending a probe field resonant with the cavity mode such that mechanical fluctuations modulate the field phase. From the measurement of the probe field phase it is possible to determine the displacement spectrum which is given by
\begin{eqnarray}
S_{qq}(\omega)\!=\!\!\!\!\int\!\!\!\dd\omega' \frac{\pq{\av{\delta\tilde q\pt{\omega}\ \delta\tilde q\pt{\omega'}}+\av{\delta\tilde q\pt{-\omega}\ \delta\tilde q\pt{\omega'}}}}{2}\,.
\end{eqnarray}
The expression for this spectrum can be decomposed in a sum of three terms as
\begin{eqnarray}
S_{qq}(\omega)=\abs{\tcmoe(\omega)}^2\ \pq{S_{\rm th}+S_{\rm rp}^{\kappa}(\omega)+S_{\rm rp}^{\rm fb}(\omega)}\,,
\end{eqnarray}
where the first is due to thermal noise and the other two are due to radiation pressure noise. They are explicitly given by
\begin{eqnarray}
S_{\rm th}&=&\gamma_{\rm m}\pt{2\,n_{\rm m}^{\rm th}+1}\,,
		\nn\\
S_{\rm rp}^{\kappa}(\omega)&=&G^2\ \kappa\ \pq{ \abs{\tcce(\omega)}^2+ \abs{\tcce(-\omega)}^2 }\,,
		\nn\\
S_{\rm rp}^{\rm fb}(\omega)&=&\frac{G^2}{2}\
\frac{\abs{\tilde\chi_{\rm fb}(\omega)}^2}{\eta\ \kappa'}
\abs{\tcce(\omega)\,\e^{-\im\theta_\Delta}+[\tcce(-\omega)]^*\,\e^{\im\theta_\Delta}}^2
\nn\\&&\hspace{-1cm}
-\frac{G^2}{2} \ [\tilde\chi_{\rm fb}(\omega)]^*\pg{
\tcce(\omega)+[\tcce(-\omega)]^*
}
\nn\\&&\hspace{1cm}\times
\pg{\tcce(-\omega)\,\e^{-\im\theta_\Delta}+[\tcce(\omega)]^*\,\e^{\im\theta_\Delta}}
\nn\\&&\hspace{-1cm}
 -\frac{G^2}{2} \ \tilde\chi_{\rm fb}(\omega)\pg{
\tcce(-\omega)+ [\tcce(\omega)]^*
 }
 \nn\\&&\hspace{1cm}\times
\pg{\tcce(\omega)\,\e^{-\im\theta_\Delta}+[\tcce(-\omega)]^*\,\e^{\im\theta_\Delta}}\ .
\nn\\
\end{eqnarray}
where $S_{\rm rp}^{\kappa}(\omega)$ is proportional to the standard radiation pressure term 
in a cavity with modified susceptibility $\tcce(\omega)$, while the term $S_{\rm rp}^{\rm fb}(\omega)$ is given by the noise term in the feedback response function $\delta\tilde\Phi_n$ [see Eqs.~\rp{noise_n} and \rp{Phi_n}].

Let us consider the term $S_{\rm rp}^{\rm fb}(\omega)$ more closely. Since we operate the feedback close to instability, where the effective cavity susceptibility $\tcce(\omega)$ has a very narrow linewidth $\kappa_{\rm eff}$, so that it is relevant only for a relatively narrow frequency range around the cavity resonance (i.e. it is peaked at $\omega=\Delta_{\rm eff}$ over a bandwidth of the order of $\kappa_{\rm eff}\ll\Delta_{\rm eff}$), we can approximate $\tcce(\omega)\ \tcce(-\omega)\sim0$, and in turn write $S_{\rm rp}^{\rm fb}(\omega)$ as
\begin{eqnarray}
	S_{\rm rp}^{\rm fb}(\omega)\!\simeq\!
	 G^2\!\pq{\ZZ_0(\omega)\, \abs{\tcce(\omega)}^2\!+\!\ZZ_0(-\omega)\,\abs{[\tcce(-\omega)]}^2}
\end{eqnarray}
with
\begin{eqnarray}
\ZZ_0(\omega)=\frac{\abs{\tilde\chi_{\rm fb}(\omega)}^2}{2\,\eta\ \kappa'} -{\rm Re}\pq{
 \tilde\chi_{\rm fb}(\omega)\ \e^{-\im\theta_\Delta}}\,,
\end{eqnarray}
which changes slowly over the range of frequencies around the mechanical mode, and can be approximated by its value close to the cavity resonance $\ZZ_0(\omega)\sim\ZZ_0(\omega)|_{\omega \sim \Delta}$.
According to Eq.~\rp{chi_fb} it can be expressed in terms of the effective cavity decay rate and detuning.

It is useful to decompose the position spectrum as
\begin{eqnarray}
S_{qq}(\omega)\simeq\abs{\tcmoe(\omega)}^2\ \pq{S_{\rm th}+S_{\rm rp}^{\rm eff}(\omega)+S_{\rm fb}(\omega)}\,,
\end{eqnarray}
where the first two terms account for the standard spectrum (with no feedback) for an optomechanical system, but with cavity decay rate $\kappa_{\rm eff}$, such that
\begin{eqnarray}
	S_{\rm rp}^{\rm eff}(\omega)&=&G^2\ \kappa_{\rm eff}\ \pq{\abs{\tcce(\omega)}^2+\abs{\tcce(-\omega)}^2}\,.
\end{eqnarray}
The last term can be interpreted as additional noise due to the feedback
\begin{eqnarray}
	S_{\rm fb}(\omega)\!&=&\!{S^{\rm fb}_{\rm rp}}(\omega)+S^{\kappa}_{\rm rp}(\omega)-S_{\rm rp}^{\rm eff}(\omega)
				\nn\\\!&=&\!
		 G^2\!\pq{\ZZ(\omega)\, \abs{\tcce(\omega)}^2+\ZZ(-\omega)\,\abs{[\tcce(-\omega)]}^2}\! ,
\end{eqnarray}
with
\begin{eqnarray}
\ZZ(\omega)&=&\ZZ_0(\omega)+\kappa-\kappa_{\rm eff}\,,
\end{eqnarray}
which changes slowly over the range of frequencies of interest, and can be approximated with its value at the cavity frequency $\ZZ(\omega)|_{\omega \sim \Delta}\equiv \ZZ^\Delta$ (see Eq.~\rp{chi_fb}) as
\begin{eqnarray}
 	\ZZ(\omega)\sim
	\ZZ^\Delta
	\nn= \frac{\pt{\Delta-\Delta_{\rm eff}}^2+\pt{\kappa_{\rm eff}-\kappa}^2}{2\,\eta\,\kappa'}\ .
\end{eqnarray}
Correspondingly
\begin{eqnarray}
S_{\rm fb}(\omega)&\sim& G^2\ \ZZ^\Delta\,\pq{ \abs{\tcce(\omega)}^2+\abs{[\tcce(-\omega)]}^2
} \ ,
\end{eqnarray}
which has the same form of the radiation pressure term, except for the factor $\ZZ^\Delta$ replacing $\kappa_{\rm eff}$. Since $\ZZ^\Delta \gg \kappa_{\rm eff}$ in the parameter range relevant to our experiment, this latter feedback noise
provides a non-negligible contribution to the overall spectrum (see Fig. 3 of the main text)
as opposed to radiation pressure.

\subsection{Steady state phonon number}

In the resolved sideband limit, that is small cavity decay rate $\kappa_{\rm eff}\ll\Delta_{\rm eff}\sim\omega_m$ and small optomechanical coupling $G\ll\omega_{\rm m}$, the steady state average number of mechanical excitations $n_{\rm m}$ can be evaluated in terms of the integral of the spectrum $S_{qq}(\omega)$~\cite{Genes2008sm}:
\begin{equation}
	n_{\rm m}+1/2 \simeq
   		\av{\delta q^2} \simeq \av{\delta p^2}
   		\simeq\frac{1}{2\,\pi}\int_{-\infty}^\infty\dd\omega\ S_{qq}(\omega).
\end{equation}
The integral can be computed by means of the analytical expressions reported in~\cite{Genes2008sm}. In order to be able to apply those formulas in the present case we have to consider another decomposition of the position spectrum. Specifically, we find that $S_{qq}(\omega)$ is proportional to the spectrum for a standard optomechanical system (with no feedback) with cavity decay rate $\kappa_{\rm eff}$ and a modified temperature, that is
\begin{eqnarray}
S_{qq}(\omega)=S'_{qq}(\omega)\ \frac{\kappa_{\rm eff}+\ZZ^\Delta}{\kappa_{\rm eff}},
\end{eqnarray}
where
\begin{eqnarray}
S'_{qq}(\omega)&\sim&
\abs{\tcmoe(\omega)}^2\ \pq{S'_{\rm th}+S_{\rm rp}^{\rm eff}(\omega)},
\end{eqnarray}
with
\begin{equation}
S'_{\rm th}=S_{\rm th}\ \frac{\kappa_{\rm eff}}{\kappa_{\rm eff}+\ZZ^\Delta} =\gamma_{\rm m}\pt{2\,n'_{\rm m}+1},
\end{equation}
and
\begin{eqnarray}
n'_{\rm m}=\frac{2\,n_{\rm m}^{\rm th}\ \kappa_{\rm eff}-\ZZ^\Delta}{2\,\pq{\kappa_{\rm eff}+\ZZ^\Delta}}\ .
\end{eqnarray}
Now the integral of $S'_{qq}(\omega)$ is readily obtained by straightforward application of the formulas reported in~\cite{Genes2008sm}. Hence we find, in the limit of small cavity decay rate $\kappa_{\rm eff}\ll\Delta_{\rm eff}\sim\omega_{\rm m}$ and when $G\ll\omega_{\rm m}$, that
\begin{eqnarray}
	\av{\delta q^2}&\sim&
		\frac{\kappa_{\rm eff}+\ZZ^\Delta}{\kappa_{\rm eff}\,\pt{\gamma_{\rm m}+\Gamma_\mathrm{eff}}}\pq{\frac{A_++A_-}{2}
			+\gamma_{\rm m}\,n'_{\rm m}\pt{1+\frac{\Gamma_\mathrm{eff}}{2\,\kappa_{\rm eff}}}},
\end{eqnarray}
with
\begin{eqnarray}
A_\pm&=&\frac{G^2\,\kappa_{\rm eff}}{\kappa_{\rm eff}^2+\pt{\Delta_{\rm eff}\pm\omega_{\rm m}}^2},\\
\Gamma_\mathrm{eff}&=&A_--A_+.
\end{eqnarray}
In particular, for high temperature, small mechanical decay rate  $\gamma_{\rm m}\ll (\Gamma_\mathrm{eff},\kappa_{\rm eff})$,  and resolved sideband limit such that  $\Gamma_\mathrm{eff}\sim A_-\gg A_+$, as in our case, the average number of mechanical excitations is
\begin{eqnarray}
	n_{\rm m}&\sim&
	n_{\rm m}^{\rm th,eff}\ \frac{\gamma_{\rm m}}{\Gamma_\mathrm{eff}}\pt{1+
		\frac{\Gamma_\mathrm{eff}}{2\,\kappa_{\rm eff}}}\,,
\end{eqnarray}
which is equal to the result for a standard optomechanical system (with no feedback), but with cavity decay rate $\kappa_{\rm eff}$, and  in a higher temperature reservoir
\begin{eqnarray}
	n_{\rm m}^{\rm th,eff} \sim
	n_{\rm m}^{\rm th}+ n_{\rm m}^{\rm eff}\ ,
\end{eqnarray}
with
\begin{eqnarray}
	n_{\rm m}^{\rm eff} =  \frac{\ZZ^\Delta\ \Gamma_\mathrm{eff}}{\gamma_{\rm m}\pt{2\,\kappa_{\rm eff}+\Gamma_\mathrm{eff}}}\ .
\end{eqnarray}

\subsection{Optomechanically induced transparency in the presence of feedback}
We now focus on the combined effect of the mechanical resonator and of the feedback loop on cavity transmission. To be more specific,
we consider the response of the system to an additional seed field injected from the input mirror, and we study how it is transmitted through the cavity. In this way we can study the effect of feedback-controlled light on optomechanically induced transparency \cite{Weis2010sm,Safavi-Naeini2011sm,Karuza2013sm}.

We are interested in the spectrum of the cavity amplitude fluctuations $\delta\tilde a+\delta\tilde a\da$ at the seed frequency, which is measured by direct photodetection of the output field $a_{\rm out}$ when the seed amplitude is much smaller than that of the pump, but still sufficiently large for the effect of the input noise operators to be negligible.
In fact, we find (neglecting terms at second order in the field fluctuations)
\begin{eqnarray}
a_{\rm out}^{\dag}\ a_{\rm out}\sim 2\,\kappa'\pq{ \alpha_s^2+\as\pt{\delta a+\delta a\da} }-\sqrt{2\,\kappa'}\,\as\pt{\ainp+a_{\rm in}^{\prime\,\dag}}\ .
\nn\\
\end{eqnarray}
From Eq.~\rp{qleom} we find that the amplitude fluctuations of the cavity field are described by the operator
\begin{eqnarray}
\delta\tilde a+\delta\tilde a\da&=&\frac{\tcmoe(\omega)}{\tcm(\omega)}\lpg{
\tcce(\omega)\ \tilde n+[\tcce(-\omega)]^*\ \tilde n\da
}\nn\\&&\rpg{
+\ii\,G\,\pq{ \tcce(\omega)- [\tcce(-\omega)]^*}\,\tcm(\omega)\ \tilde\xi.
}
\end{eqnarray}
In order to compute the transmission of the seed field we can include the seed amplitude at frequency $\nu$ in the noise operator $\tilde n\to \tilde n+\sqrt{2\,\kappa_0}\, \ee^{-\ii\,\theta_{\Delta}}\,\alpha_{seed}\ \delta(\omega-\nu)$.
Thereby we find that, neglecting all the noise terms, the transmitted field close to the cavity resonance ($\nu\sim\Delta$) is
\begin{equation}
\tilde a_{\rm out}^{\dag}\ \tilde a_{\rm out} \sim  2\,\kappa'\,\as\pt{\delta\tilde a+\delta\tilde a\da}
\sim \tilde t(\nu)\ \alpha_\mathrm{seed},
\end{equation}
with the transmission coefficient given by
\begin{eqnarray}
\tilde t(\omega)=2\,\kappa'\sqrt{2\,\ko}\,\e^{-\im\theta_\Delta}\
\tcce(\omega)\ \frac{\tcmoe(\omega)}{\tcm(\omega)},
\end{eqnarray}
where
\begin{eqnarray}
\frac{
\tcmoe(\omega)}{\tcm(\omega)}&\sim&\frac{[\tcm(\omega)]^{-1}}
{
[\tcm(\omega)]^{-1}-\ii\,G^2\ \tcce(\omega)
}
\nn\\
&=&
\frac{
\omega_{\rm m}^2-\omega^2-\ii\,\omega\,\gamma_{\rm m}
}{
\omega_{\rm m}^2-\omega^2-\ii\,\omega\,\gamma_{\rm m}
-\ii\,\omega_{\rm m}\,G^2\ \tcce(\omega)
}\ .
\end{eqnarray}

\begin{figure}[th!]
\includegraphics[width=0.49\textwidth]{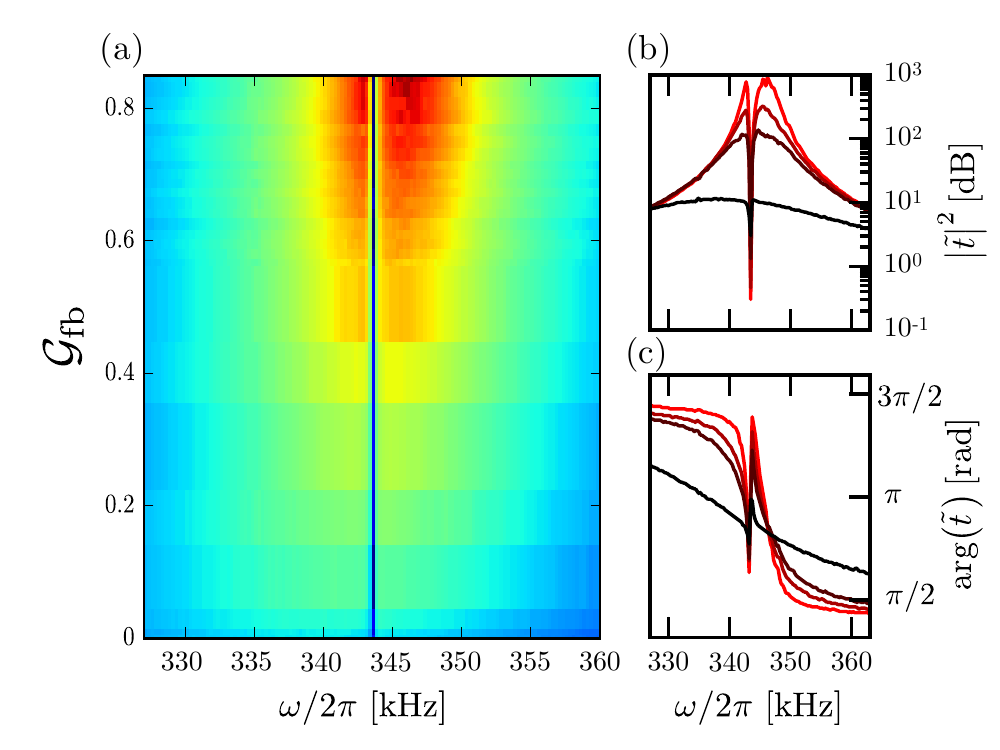}
\caption{Optomechanically induced transparency (OMIT) in the presence of feedback. (a) Color plot of the modulus square of the cavity transmission $|t(\omega)|^2$, as a function of frequency (horizontal scale) and feedback gain normalized as defined in the Letter (vertical scale). The measurement is performed by injecting a seed on the in-loop cavity mode. The value of $|t(\omega)|^2$ is increasing from blue to red. At the mechanical resonance frequency, the seed is no more transmitted. The interference between the seed and the sideband created by the mechanical mode on the red--detuned beam determines a destructive interference, that is the OMIT phenomenon. Its width is determined by the optomechanical coupling, which is fixed, while the cavity decay rate $\kappa_{\rm eff}$ is modified by the feedback loop and decreases for increasing feedback gain towards the instability. In (b) and (c) we plot, respectively, the magnitude square and phase of the cavity transmission for different fixed feedback gain increasing from dark to light red. The black trace is obtained with no feedback.}
\label{fig:omit}
\end{figure}

The transmission spectrum is then given by
\begin{equation}
S_t(\omega)=\abs{\tilde t(\omega)}^2=8\, \kappa_0\,{\kappa'}^2\ \abs{\tcce(\omega)}^2\ \abs{\frac{\tcmoe(\omega)}{\tcm(\omega)}}^2,
\end{equation}
and can be
expressed as a standard Fano profile~\cite{Fano1961sm,Lounis1992sm}, which describes interference phenomena, as
\begin{eqnarray}
S_t(\omega)&=&
8\, \kappa_0\,{\kappa'}^2\ \abs{\tcce(\omega)}^2\
\pq{\frac{(\epsilon+q)^2}{\epsilon^2+1}+\rho},\label{Fano}
\end{eqnarray}
where
\begin{eqnarray}
\epsilon&=&\frac{
{\omega^2-\omega_{\rm m}^2}-{\omega_{\rm m}}\,G^2\,{\rm Im}\pq{\tcce(\omega)}
}{
\gamma_{\rm m}\,{\omega}+{\omega_{\rm m}}\,G^2\,{\rm Re}\pq{\tcce(\omega)}
},
\nn\\
q&=&\frac{
{\omega_{\rm m}}\,G^2\,{\rm Im}\pq{\tcce(\omega)}
}{
\gamma_{\rm m}\,{\omega}+{\omega_{\rm m}}\,G^2\,{\rm Re}\pq{\tcce(\omega)}
},
\nn\\
\rho&=&\frac{\gamma_{\rm m}^2\,{\omega^2}}{
\abs{
\omega_{\rm m}^2-\omega^2-\ii\,\omega\,\gamma_{\rm m}
-\ii\,\omega_{\rm m}\,G^2\, \tcce(\omega)}^2
}\ .
\end{eqnarray}
Notice from Eq. (\ref{Fano}) that the Fano profile is determined by the {\it effective} cavity susceptibility modified by the feedback loop, $\tcce(\omega)$. Destructive interference is observed when $\epsilon=-q$, that is when $\omega=\omega_{\rm m}$, and $\rho$ represents an additional small term proportional to the mechanical damping rate $\gamma_{\rm m}$ which prevents perfect destructive interference.
When $q=0$ at the interference point $\omega=\omega_{\rm m}$ (that is when ${\rm Im}[{\tcce(\omega_{\rm m})}]=0$ , i.e. $\Delta_{\rm eff}=\omega_{\rm m}$), the spectrum is symmetric with a dip in the middle. Instead, the spectrum is asymmetric when $q\neq0$ for $\omega=\omega_{\rm m}$.
The behaviour of the OMIT, the same as in standard optomechanical systems, but with the cavity response modified by the feedback, is experimentally verified in~\figurename~\ref{fig:omit}.

\section{Fine--gain attenuator calibration}
\begin{figure}[th!]
\includegraphics[width=0.49\textwidth]{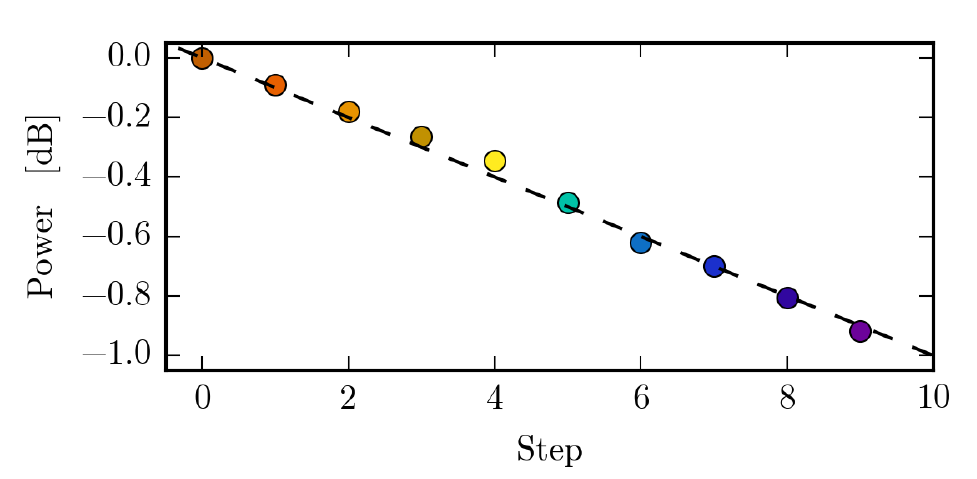}
\caption{Calibration of the fine gain attenuator.}
\label{fig:GainCal}
\end{figure}

To explore deeply the instability region, and to reach the normal--mode  splitting regime, we have realised a circuit for a fine  tuning of the feedback gain. An overall attenuation of $\SI{1}{\decibel}$ is divided in ten steps, each determined by three appropriate resistors in Pi--configuration. The calibration of the feedback gain steps, which are used for the evaluation of the feedback gain in the experimental analysis, is reported in~\figurename~\ref{fig:GainCal}.

\end{document}